\documentclass[prd,showpacs,preprint,11pt]{revtex4}
\usepackage{graphicx}
\newcommand{\rd}{\mathrm{d}}
\usepackage{amsfonts}
\usepackage{amsmath}
\usepackage{amssymb, graphics}

\begin{document}
\title{Singularity and  entropy in the bulk viscosity dark energy model }
\author{Xin-He Meng$^{1,2}$}
\email{xhm@nankai.edu.cn}
\author{Xu Dou$^{1}$}
\email{dowxdou@gmail.com}

\affiliation{$^{1}$Department of physics, Nankai University,
Tianjin, 300071, China\\$^{2}$Kavli Institute of Theroretical
Physics China,\\CAS, Beijing 100190, China}
\begin{abstract}
In this paper bulk viscosity is introduced to describe the effects
of cosmic non-perfect fluid on the cosmos evolution and to build the
unified dark energy (DE) with (dark) matter models. Also we derive a
general relation between the bulk viscosity form and Hubble
parameter that can provide a procedure for the viscosity DE model
building. Especially, a redshift dependent viscosity parameter
$\zeta\propto\lambda_{0}+\lambda_{1}(1+z)^{n}$ proposed in the
previous work by X.H.Meng and X.Dou in 2009\cite{md} is investigated
extensively in this present work. Further more we use the recently
released supernova dataset (the Constitution dataset) to constrain
the model parameters. In order to differentiate the proposed
concrete dark energy models from the well known $\Lambda$CDM model,
statefinder diagnostic method is applied to this bulk viscosity
model, as a complementary to the $Om$ parameter diagnostic and the
deceleration parameter analysis performed by us before. The DE model
evolution behavior and tendency are shown in the plane of the
statefinder diagnostic parameter pair \{$r,s$\} where the fixed
point represents the $\Lambda$CDM model. The possible singularity
property in this bulk viscosity cosmology is also discussed to which
we can conclude that in the different parameter regions chosen
properly, this concrete viscosity DE model can have various late
evolution behaviors and the late time singularity could be avoided.
We also calculate the cosmic entropy in the bulk viscosity dark
energy frame, and find that the total entropy in the viscosity DE
model increases monotonously with respect to the scale factor
evolution, thus this monotonous increasing property can indicate an
arrow of time in the universe evolution, though the quantum version
of the arrow of time is still puzzling.

\end{abstract}


\newpage

\maketitle
\section{Introduction}

Type Ia supernova and other astrophysics observations together
indicate that our universe is accelerating now \cite{sn1}. Different
models are proposed to try to describe or understand this
surprisingly exotic phenomenon. If we make the assumption that
general relativity is still correct to the scale of cosmos, an
effective term contributes to negative pressure should be added to
the right hand side of Einstein's field equation in the general
theory of relativity to explain the recent stage speed-up of our
observational universe expansion. This is the basic idea to the so
called dark energy concept. So far for the ten years' old DE
phenomena there are many both theoretical and observational attempts
to understand the mechanism. The introduction of the cosmological
constant corresponds to a negative pressure fluid with a specially
constant density all the time which has been playing a key role in
the universe evolution by uniformly distributed over the whole
cosmic space-time and media, but the cosmological constant existence
raises serious fundamental physics problem, the so-called
cosmological constant problem for both new (coincidence) and old (so
tiny) ones. Another class of models try to modify the traditional
Einstein's general theory of relativity to the large cosmology
scale, by arguing that the recently appearing acceleration phase of
the unverse expansion comes from the break down of general
relativity in cosmic scale. The so-called $f(R)$ gravity, to mention
one for example, which generalizes the Hilbert-Einstein action is
categorized in this class \cite{mw1}. (For more details and
references, you may see the recent reviews \cite{r1} \cite{r2}).

In the context of perfect fluid, some models based on fluid
mechanics method are studied extensively, such as the Chaplygin gas
and generalized Chaplygin gas models which modify the equation of
state, and barotropic fluids dark Energy \cite{ls}. Also for the
purpose to consider more realistic situation in the evolution of the
universe, the concept of viscosity is introduced into the
investigation of the cosmos evolution from fluid mechanics \cite{v2}
\cite{v3} \cite{v7} \cite{v8} \cite{v9} \cite{v10} \cite{v11}.
Earlier attempt \cite{pc} in this area even ``predicts'' the late
acceleration of the universe expansion. The dissipative effect in
the fluid is always due to shear and bulk viscosity characterized by
shear viscosity parameter $\eta$ and bulk viscosity parameter
$\zeta$. In the study of cosmology, the shear viscosity disappears
in the Friedmann-Robertson-Walker's isotropic and homogeneous
framework with the largest spherical symmetry. Only bulk viscosity
could play a role in the realistic models. Its effects can be shown
from an added correction term with the minus sign to cosmic pressure
$\tilde p=p-3\zeta H$. This composite formula motives the trial on
the connection between bulk viscosity of the cosmic fluid and dark
components (energy and matter). As we understand so far that purely
gravitational probes can only provide information on a single
effective matter-energy fluid, which may consist of the dark energy
and (dark) matter composites as well as the least dominated
radiation energy at present universe, so for the dominated two
functioning dark components we can describe them as a single dark
fluid \cite{Ren}. In this paper, we discuss a concrete unified model
building by paying attention on the modification of the cosmic media
from the simple assumption as a perfect fluid to a piratical viscous
fluid encoded in the energy-stress tensor contents, that is, at the
right hand side of the Einstein's field equation. For the bulk
viscosity DE model building a remark is needed here. Considering the
observational dark energy ingredient fraction today, it takes about
$\frac{2}{3}$ of the whole matter-energy density, so the
corresponding pressure provided by the bulk viscosity correction
dominantly surpasses the remaining pressure contributions from other
cosmic energy-matters. This is obviously different from the
traditional non-equilibrium thermodynamics, in which the viscosity
contribution is only a little correction to the pressure term. Some
researches try to find a mechanism to support such a fluid behavior
by a kind of non-standard interaction introduced between the dark
matter and energy components \cite{inter1} \cite{inter2}. It is
certainly important for these attempts to find more support or hints
behind cosmic observations in the study of bulk viscosity cosmology
along this possible line.

For concrete model building, detailed form of viscosity parameter is
needed to settle down the theoretical framework. In the reference
\cite{Ren}, a scale factor dependent viscosity is proposed, which is
different from the only density dependent form
\begin{equation}
\zeta=\zeta_{0}+\zeta_{1}\frac{\dot a}{a}+\zeta_{2}\frac{\ddot
a}{\dot a}
\end{equation}
It could be shown that it is equivalent to a modified equation of
state(EOS)
\begin{equation}
p=(\gamma-1)\rho+p_{0}+w_{H}H+w_{H2}H^{2}+w_{dH}\dot H.
\end{equation}
And other interesting physical properties have been investigated in
that kind of models \cite{rm} \cite{con}. In this present paper, by
largely extending its contents we will continue the study of a
redshift dependent viscosity form
$\zeta\propto\lambda_{0}+\lambda_{1}(1+z)^{n}$ as proposed in the
previous work \cite{md}.  We concentrate on the late time
singularity discussion and its entropy expression of the bulk
viscosity DE model. For the phantom dark energy model, there exists
a cosmic singularity in the future cosmos evolution, that is, the
so-called cosmic ``doomsday''. As shown below, we could see that
this viscosity DE model represents different evolution behaviors,
and the future cosmic singularity will disappear under some proper
parameter ranges selection. From this view of point, the model has
the flexibility to produce either quintessence or phantom
properties, that is, we can easily achieve its EoS either larger or
less than characteristic -1.

An important tool for investigating dark energy model characters
nowadays is by the introduction of some geometry quantities, for
instance the statefinder diagnostic parameters \cite{sf}, which are
quantities dependent on high order derivatives of the scale factor,
such as to the $\dddot a$. The usual statefinder parameter pair
\{$r,s$\} of the model concerned is calculated explicitly to
demonstrate the viscosity DE model behaviors. In our previous work,
deceleration parameter and $Om$ diagnostic parameter \cite{om} are
performed to show the properties. We have found that in statefinder
pair plane, our model and $\Lambda$CDM model could be easily
discriminated in some redshift ranges. We prospect that increasing
the quality and quantity of measurement data will enhance our
ability to make accurate discrimination of different current
cosmology models and rule out some. At the same time, new diagnostic
methods merit further investigations, especially diagnostic quantity
in the higher order perturbation level.

The paper is organized as follows: in the second section, some
general features and remarks about viscosity dark energy model are
given, and we further discuss the redshift dependent model. In the
third section, we give the singularity discussion of our model. Some
solutions are given there. In Sec. \textbf{IV}, we calculate entropy
of this viscosity model. Finally, conclusion and discussions are
presented. We leave the data fitting procedure in the appendix.

\section{Viscosity Dark Energy Model}
As well-known so far, the general theory of relativity for gravity
and standard model for particle physics are very successful to
describe the universe phenomena before the astrophysics exotic
behaviors discovered, such as the roughly flat rotation curves for
the spiral galaxies at large distance and the speed-up for the
universe expansion at current stage evolution that are now commonly
attributed as the cosmic "dark" physics evidences. The cosmic dark
sector, often divided as mainly the dark energy and dark matter
parts, consists of about $95\%$ of the total cosmic ingredients and
we know that purely gravitational probes are blind to the two main
"flavors", that is the dark matter and dark energy can not be
separated clearly. So at present evolution stage for our physical
universe, a single unified dark fluid follows to describe the main
cosmic media. In the Friedmann Robertson Walker metric the cosmic
fluid is usually described by its homogenous and isotropic density
$\rho$ as well as the pressure $p$, that is, the pressure $p$ can be
divided into two additive parts (dark energy and matter) at least:
\begin{equation}\label{eq1}
p=p_{m}+p_{de}.
\end{equation}
Here we assume that other cosmic components, such as radiation and
curvature contributions, are negligible as they may play
un-important roles in our present description to the late universe
evolution. It may be a simplifying view to treat cosmic fluid
dividedly, but we consider that it is a more practical way for
modeling dark energy and dark matter in a unified single fluid  for
the study of the late universe evolution. Also in our concrete
viscosity DE model building, the usual assumption that cosmic fluid
simply is perfect is not kept. Instead, we assume that the cosmic
fluid for current universe evolution is better described by a single
non-perfect fluid encoded with bulk viscosity effects. Therefor the
modified pressure with viscosity term can be written directly as:
\begin{equation}
\tilde p=(\gamma-1)p-3\zeta H,
\end{equation}
where $\zeta$ is the bulk viscosity parameter. It is useful to keep
in mind that the relation between single fluid and multi-fluid
framework, that is, in the multi-fluid case, the dark energy
pressure $p_{de}$ comes mainly from a bulk viscosity term added to
the perfect fluid. This treatment could also be regarded as an
effective method which revises the dark energy equation of state
(EoS). If we set thermodynamics index $\gamma=1$ in this paper, the
cosmic pressure is mainly due to the viscous effects. Using the
Friedman equation, we can get the revised equation of state for the
relatively complex cosmic fluid:
\begin{equation}\label{eq3}
\tilde{p}=-3\kappa \zeta \sqrt{\rho}
\end{equation}
where in the equation above $\kappa^2=\frac{8\pi G}{3}$.\\
If we define the effective EoS as usual $\omega=p/\rho$ ,then in
this case:
\begin{equation}
\omega=-3\kappa\frac{\zeta}{\sqrt{\rho}}
\end{equation}
With the aim to be consistence with current astrophysics observation
results, the value of equation of state today should be
\begin{equation}
\omega|_{0}\sim-1
\end{equation}
where the sub script zero indicates today's value, so that,
\begin{equation}
\zeta\sim\frac{H^{2}}{3\kappa^{2}}
\end{equation}
If we fix the equation of state, that is, $\omega$ is a constant
$\omega_{0}$, we can have an interesting relation from eq.(6)
\begin{equation}
\rho=\frac{9\kappa^{2}\zeta^{2}}{\omega_{0}^{2}}
\end{equation}
Three aspects on the viscosity DE model building are detailed below.
\subsection{A General function for the Hubble parameter with viscosity contribution}
For consistent with observation data, especially supernova data, we
need calculate the integration of Hubble parameter, such as quantity
like $\int \frac{1}{H(z)}\rd z$, so to obtain more knowledge on
$H(z)$ is certainly very useful. With the improved data quality and
quantity of direct $H(z)$ parameter observations (like the Hubble
Space Telescope project now running in the sky) we can use this
better constrained Hubble rate in the fitting procedures to get more
information on our observational universe evolution.

The conservation equation with a viscosity term is:
\begin{equation}
\dot\rho+3H\rho=9\zeta H^{2}.
\end{equation}
With the Friedman equation, we can rewrite the  conservation
equation in terms of a Hubble parameter:
\begin{equation}
\dot H+\frac{3}{2}H^{2}=\frac{3}{2}\zeta H.
\end{equation}
Using the relation $\rd t=\frac{1}{aH}\rd a$, we can write the above
formula as a differential equation to the scale factor:
\begin{equation}
\frac{\rd H}{\rd a}+\frac{3}{2a}H=\frac{3\zeta}{2a}.
\end{equation}
Its form solution is thus obtained as a general integral function of
the scale factor:
\begin{equation}\label{H}
H=C_{1}a^{-3/2}+\Big[\int\frac{3\zeta}{2a}\mathrm{exp}\big(\int\frac{3}{2a}\rd
a\big)\rd a\Big]\mathrm{exp}\big(-\int\frac{3}{2a}\rd a\big).
\end{equation}
where the coefficient $C_1$ is an integral constant. For a different
viscosity form $\zeta(a)$ given out, we could derive a concrete
Hubble parameter expression accordingly. So we have obtained a
general way to build the bulk viscosity dark energy models. We note
that the emergent of the first term in eq. (\ref{H}) which looks
like the matter dominated contribution. By making the single fluid
assumption, that is, the concrete ingredient of cosmic density is
not specified, the $a^{-3/2}$ term naturally arises besides others,
and contributes to a $a^{-3}$ term in the function of $H^{2}$. Best
fitting results with available observational data in previous work
\cite{md} is consistence with identifying the coefficient of this
term as dark matter mass ratio $\Omega_{m}^{1/2}$.

\subsection{The Redshift Dependent Viscosity}
When the bulk viscosity parameter is specified, we could discuss
cosmology with the unified dark energy evolution behaviors. Variable
viscosity parameter like the density dependent model
$\zeta(\rho)=\alpha\rho^{m}$ has been investigated extensively as in
ref. \cite{rho}. In the previous work, we have proposed a new
analytical form of viscosity parameter in the flat FRW space-time:
\begin{equation}
9\lambda=\lambda_{0}+\lambda_{1}(1+z)^{n},
\end{equation}
where the viscosity is re-written as
$\lambda=H_{0}\zeta/\rho_{cr0}$, and $z$ is the redshift. The
exponent parameter n is to be best fitted by observational data
sets, so as the $\lambda_{0}$ and $\lambda_{1}$  two arbitrary
constants.From Friedmann equation containing the viscosity term we
have:
\begin{equation}
\frac{\ddot a}{a}=-\frac{\kappa}{2}(\rho+3\tilde{p}),
\end{equation}
that is exactly in detail
\begin{equation}\label{eq13} \frac{\ddot
a}{a}=-\frac{\kappa}{2}\big(\frac{\dot
a}{a}\big)^2+\frac{3}{2}\kappa^2\zeta\frac{\dot a}{a}.
\end{equation}
We could derive the function of Hubble parameter $H$ evolution with
the redshift as:
\begin{equation}
\frac{H}{H_{0}}=\lambda_{2}(1+z)^{1.5}-\frac{\lambda_{1}}{2n-3}(1+z)^{n}+\frac{\lambda_{0}}{3},
\end{equation}
The coefficient $\lambda_{2}$ is an integration constant and
constrained self-consistently by the above relation when taking z=0.
We use this function of Hubble parameter by fitting the latest
released Constitution supernova data sets to get these free constant
parameters. By the numerical fitting processes with statistic
analysis that is compiled in the appendix of this work, and choose
the parameter $n=-1$ as the best value, we could get a result closed
to the $\Lambda$CDM model when comparing the deceleration parameter
derived from different models with various n values. In the FIG. 1,
the theoretical distance modulus curve of this redshift dependent
viscosity DE model is plotted, which gives an acceptable fitting
results with the latest released supernova data sets.

\begin{figure}
\includegraphics{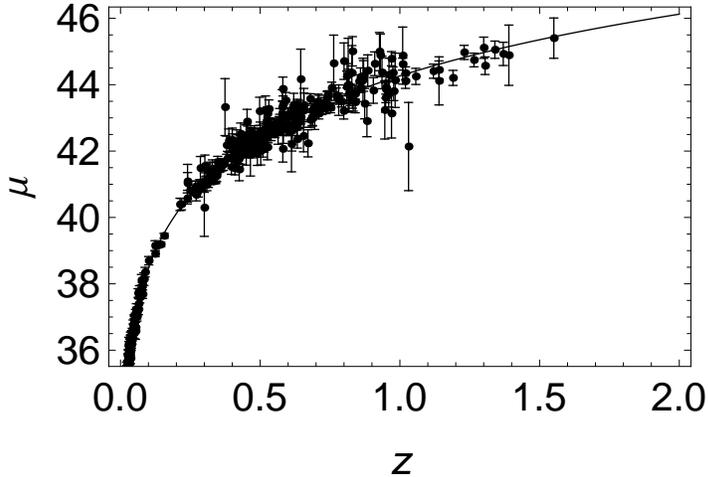}
\caption{The relation between distance modulus and redshift. The
solid line corresponds to the theoretical curve of our viscosity DE
model and the dots with error bar are the observational data.}
\end{figure}

\subsection{Statefinder diagnostic analysis}
Two statefinder parameters are defined in terms of the scale factor
and Hubble parameter by
\begin{equation}\label{1}
r=\frac{\dddot{a}}{aH^{3}}, \qquad s=\frac{r-1}{3(q-\frac{1}{2})},
\end{equation}
where the deceleration parameter $q$ is defined as
$q=-\frac{\ddot{a}}{aH^{2}}$.

If we have already obtained the function form of Hubble parameter
$H(z)$, it is more convenient to express the diagnostic statefinder
parameter pair \{${r,s}$\} in terms of $H(z)$, therefore $r(z)$ and
$q(z)$ become
\begin{equation}
r=(1+z)q^{\prime}-q[1-2(1+z)\frac{H^{\prime}}{H}],
\end{equation}
\begin{equation}
q=(1+z)\frac{H^{\prime}}{H}-1,
\end{equation}
where the prime represents derivative with respect to the redshift
$z$.

Functional relation between the diagnostic parameters $r$ and $s$ is
displayed above. In the $r-s$ plane, the fixed dot at \{${{0,1}}$\}
corresponds to the concordant $\Lambda$CDM model. Here, we only
consider the special case in which the parameter $n=-1$ for the
viscosity DE model and other parameters $\lambda_{1}$, and
$\lambda_{2}$ have been best fitted in this case.

\begin{figure}
\includegraphics{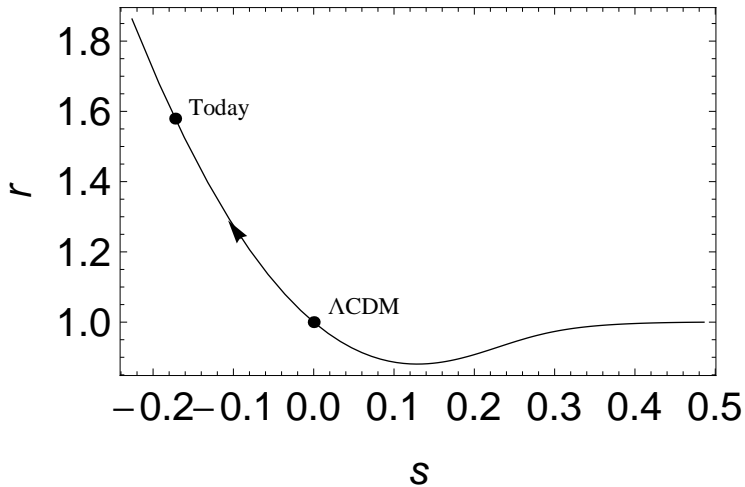}
\caption{The universe evolution is diagnostic by the evolution
trajectory in the $r-s$ plane for the $n=-1$ condition. The fixed
point at (0,1) corresponds to the $\Lambda$CDM model.}
\end{figure}

The universe evolution curve by the statefinder dialogistic
parameter passes the point which corresponds to the $\Lambda$CDM
model. For our model with the viscosity effects in the best fitting
case, this point is just at when the universe evolves at redshift
$z\simeq0.447$  where the statefinder diagnostic parameters are
\{$s=0,r=1$\} by numerical calculations. We conclude that the
statefinder diagnostic analysis shows model degeneracies (the
$\Lambda$CDM model and our viscosity DE model) at this special
point, but it could be easily to discriminate between viscosity DE
models and the $\Lambda$CDM model elsewhere. The statefinder
diagnostic parameter values today with the redshift z=0 correspond
to the point with $r=1.579$ and $s=-0.173$. Its location is
represented in the figure 2, too.
\begin{center}
\begin{figure}
\includegraphics{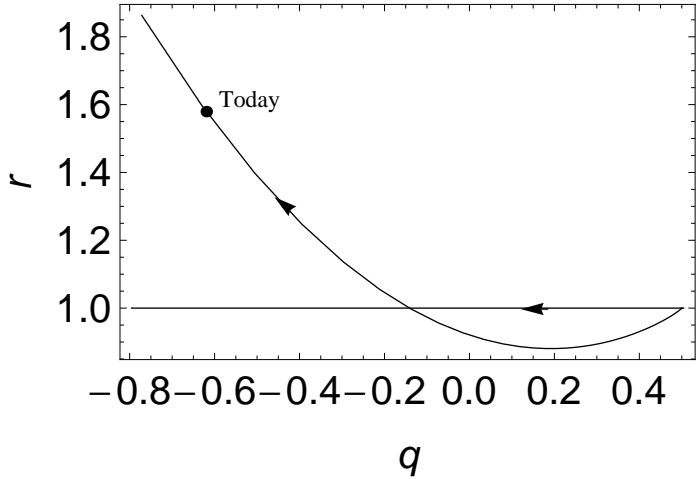}
\caption{Evolution trajectory in $r-q$ plane to demonstrate the
universe evolution in another diagram view.}
\end{figure}
\end{center}
To illustrate more details for the unified viscosity DE model
different from the $\Lambda$CDM model we also plot the parameter
r(q) function trajectory in the $r-q$ plane as shown in the figure
3. The black arrow points towards the direction the universe evolves
to and the horizonal line corresponds to the $\Lambda$CDM model
evolution. We know that the negative geometric quantity deceleration
parameter $q$ represents that the universe expansion is undergoing
an acceleration stage, hence it could be obviously shown that the
evolution tendency trajectory from the cosmic expansion deceleration
range to expansion acceleration range, which corresponds to our
universe evolution from the right to the left side along the
horizonal q axis in the $q-r$ plate.

To demonstrate further the similarities and differences between the
unified viscosity DE model and the $\Lambda$CDM model we  plot and
compare the q-z relations of the two models as shown in the figure
4, too, which provides another view to compare the two models.
\begin{figure}
\includegraphics{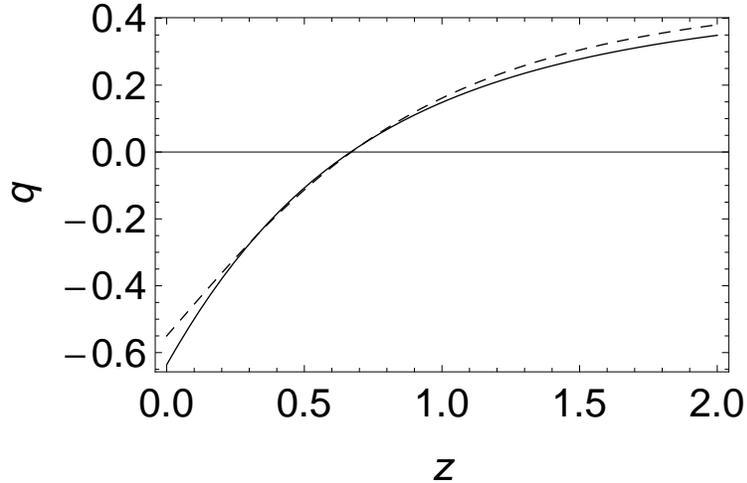}
\caption{The $q-z$ relation diagram. The dashed line corresponds to
the deceleration parameter relation with the redshfit computed from
the $\Lambda$CDM model with the model parameter $\Omega_{m}=0.3$,
while the solid line delineates the best fitted viscosity DE model
discussed here.}
\end{figure}

\section{the Scale factor and the finite future Singularity}

It is well known that the phantom dark energy models with a negative
equation of state have had exotic cosmic evolution
behaviors\cite{ps}. Cosmos evolution may be driven to a singularity
in the finite future, such as the so called Sudden, Big-rip, or
Big-brake singularities\cite{si} \cite{1}. And as for the newly
explored type of singularity named Barotropic index w-singularities
\cite{bs}, it could be led by a scale factor with a typical
expression:
\begin{equation}
a(t)=A+B\bigg(\frac{t}{t_{s}}\bigg)^{\frac{2}{3\gamma}}+C\bigg(D-\frac{t}{t_{s}}\bigg)^{n}.
\end{equation}
where the parameter $t_s$ indicates a finite future time when the
singularity will occur and the A, B, C, D, n and $\gamma$ are all
free parameters to be determined. Singularities are harmful to the
prediction power for a theory and theoretical model buildings. With
the Hubble parameter got previously, we could discuss more details
on the evolution properties, especially the future singularity of
our viscosity DE model. Since the scale factor $a(t)$ can be
expressed in terms of the redshift as $\frac{1}{1+z}$ when we set
$a(0)=1$, the Hubble parameter can thus be rewritten as:
\begin{equation}
\frac{H}{H_{0}}=\lambda_{2} a^{-1.5}-\frac{\lambda_{1}}{2n-3}
a^{-n}+\frac{\lambda_{0}}{3}.
\end{equation}\\
As we can easily see above that the unified viscosity DE model
behaves generally as the $\Lambda$CDM model in the early universe
evolution, especially when n negative, we will discuss several
special cases to the above general evolution features of eq.(22)\\
(a) the $n<0$ case\\
with the increasing of the scale factor, the first term,
proportional to $a^{-1.5}$, will decay to zero, while the n-power
term will increase and dominate in the future universe evolution.
The Hubble parameter will be approximated then as:
\begin{equation}
\frac{H}{H_{0}}\simeq-\frac{\lambda_{1}}{2n-3}a^{-n}+\frac{\lambda_{0}}{3}.
\end{equation}
For the $n=-1$ case, in which we have a better comparison with the
$\Lambda$CDM model. The differential equation for the scale factor
evolution is:
\begin{equation}
\dot a
\simeq\frac{\lambda_{1}H_{0}}{5}a^{2}+\frac{\lambda_{0}H_{0}}{3}a.
\end{equation}
We could get an approximate solution for the scale factor evolution
behaviors:
\begin{equation}
a(t)=
\frac{5\lambda_{0}}{3\lambda_{1}}\frac{\mathrm{exp}(\frac{\lambda_{0}H_{0}}{3}t-\frac{\lambda_{0}H_{0}}{3}t_{s})}{1-\mathrm{exp}(\frac{\lambda_{0}H_{0}}{3}t-\frac{\lambda_{0}H_{0}}{3}t_{s})},
\end{equation}
where $t_{s}$ is the time parameter when a future singularity
happens (when $t=t_{s}$, the
scale factor has a finite future singularity: $a\rightarrow\infty$).\\
As we all know, the scale factor $a(t)$ is a basic and crucial
quantity to our understanding of the cosmic evolution history. We
may use this approximate solution of the scale factor in the future
evolution to discuss more details of this viscosity DE model near
future singularity. The cosmic media density reads as:
\begin{equation}
\rho=\frac{3H^{2}}{\kappa}.
\end{equation}
Inserting the approximated expression of Hubble parameter above into
the above expression we can have:
\begin{equation}
\rho=\frac{3}{\kappa}\Big(\frac{\lambda_{1}H_{0}}{5}a+\frac{\lambda_{0}}{3}\Big)^{2}.
\end{equation}
We could directly see that if $a\rightarrow\infty$ as $t\rightarrow
t_{s}$, the
cosmic density will diverge, too.\\
 By the Eq.(\ref{eq3}), we could also see that the cosmic pressure $|p|\rightarrow\infty$ as
$a\rightarrow\infty$. From the EoS defined above we obtain:
\begin{equation}
\omega=-3\kappa\frac{\zeta}{\sqrt{\rho}},
\end{equation}
we have $\omega\rightarrow\omega_{s}$ where the $\omega_{s}$ is the
EoS when $a\rightarrow\infty$. In the ref. \cite{1}, authors
classify the future singularities in the
following way:\\

   (i) Type \textbf{I}(``Big Rip''): For $t\rightarrow t_{s}$,
   $a\rightarrow\infty$, $\rho\rightarrow\infty$, and
   $|p|\rightarrow\infty$,\\

   (ii) Type \textbf{II}(``Sudden''): For $t\rightarrow t_{s}$,
   $a\rightarrow a_{s}$, $\rho\rightarrow\rho_{s}$, and
   $|p|\rightarrow\infty$,\\

   (iii) Type \textbf{III}: For $t\rightarrow t_{s}$, $a\rightarrow
   a _{s}$, $\rho\rightarrow\infty$, and $|p|\rightarrow\infty$,\\

   (iv) Type \textbf{IV}: For $t\rightarrow t_{s}$, $a\rightarrow
   a_{s}$, $\rho\rightarrow 0$, $|p|\rightarrow 0$, and higher
   derivatives of $H$ diverge.\\

We could see that the model we discussed above falls basically into
the Type \textbf{I} category , that is, the so-called  ``Big Rip''
singularity. Another relevant problem to solve is the determination
of singularity time $t_{s}$. In the discussions above, the parameter
$t_{s}$ emerges as an integration constant. When $t$ increases to
some finite value, cosmos will come to the singularity, and we can
choose this value as the cosmic singularity time $t_{s}$. Some
authors \cite{2} have given an approximation value of a cosmic
``doomsday'', as $t_{s}=35$ Gyr. As for the early universes evolution in
our unified viscosity DE model it returns to the Friedmann phase or de Sitter phase.\\
(b)  The $n=0$ case\\
In this condition, the viscosity parameter will became a constant
$\zeta=\frac{\rho_{cr0}}{H_{0}}(\lambda_{0}+\lambda_{1})$. Since
$h(z=0)=1$ in the special case, it is easy to verify that $\zeta>0$.
Hence, the scale factor solution to the Friedmann equation
Eq.(\ref{eq13}) is
\begin{equation}\label{so}
a(t)=k\big[1+(1+m)\mathrm{exp}(\frac{3}{2}\kappa^2\zeta,
t)\big]^{\frac{1}{1-m}}
\end{equation}
where $k$ is a normalization factor, and $m=\frac{\kappa}{2}$.\\
And in this case, there is no the finite time singularity. A de
Sitter phase of universe will emerge as $t\rightarrow\infty$. A
parallel remark could be made here that for the $n<0$ case, we can
neglect the $\lambda_{1}(1+z)^n$ term from the viscosity
contributions. Then we have obtained a solution with the same form
as Eq. (\ref{so}). This is an inflation-like solution, but when
$t=0$, $a(init)\neq0$, there is a initial
condition uncertainty problem.\\
(c) The $n>0$ case\\
Terms about scale factor in Hubble parameter will decay and vanish,
so as $t\rightarrow\infty$, the Hubble parameter will become a
constant. Then, the term $\frac{\lambda_{0}}{3}$ will play a role as
the effective cosmological constant which dominates the late
universe evolution, therefore the universe will enter into a de
Sitter phase.

\section{Entropy and an arrow of time}
Entropy is related causality and the second law of thermodynamics is
usually regarded as the major physical manifestation of the arrow of
time, from which many interesting consequences can be derived. In
this section, we will discuss the entropy expression in our unified
viscosity DE model. Due to the non-perfect fluid property assumption
of the viscosity DE model, the entropy will change in contrast with
the case for perfect fluid models, in which
$\frac{\rd S}{\rd t}=0$ (Where we define $S$ as the entropy of the system in unit volume).\\
In Refs. \cite{3,4,5,hm}, the general formula for the entropy
expression has been given as:
\begin{equation}\label{eq17}
S^{\mu}_{;\mu}=\frac{\eta}{T}\sigma_{\mu\nu}\sigma^{\mu\nu}+\frac{\zeta}{T}\theta^2+\frac{1}{\kappa
T}Q_{\mu}Q^{\mu},
\end{equation}
where the $S^{\mu}$ is the entropy four-vector, $\eta$ the shear
viscosity, T the system temperature, $\sigma_{\mu\nu}$ the shear
tensor, $\theta$ the expansion factor, $\kappa$ the thermal
conductivity and the four-vector $Q^{\mu}$ the space-like heat flux
density. And the entropy four-vector $S^{\mu}$ is defined by:
\begin{equation}
S^{\mu}=\delta SU^{\mu}+\frac{1}{T}Q^{\mu},
\end{equation}
where $\delta S$ is the ordinary entropy per unit volume in single
fluid, and the contributions from different ingredients are not
specified. In the universe evolution system the expansion tensor
$\theta_{\mu\nu}$ is defined as
\begin{equation}
\theta_{\mu\nu}=\frac{1}{2}(U_{\mu;\alpha}h^{\alpha}_{\nu}+U_{\nu;\alpha}h^{\alpha}_{\mu}).
\end{equation}
The scalar expansion factor thus is
$\theta=\theta^{\mu}_{\nu}=U^{\mu}_{;\mu}$, especially in the FRW
background $\theta=3H$. The shear tensor is defined as
$\sigma_{\mu\nu}=\theta_{\mu\nu}-\frac{1}{3}h_{\mu\nu}\theta$, where
$h_{\mu\nu}$ is defined by $h_{\mu\nu}=g_{\mu\nu}+U_{\mu}U_{\nu}$.
Defining the four-acceleration of the fluid as $A_{\mu}=\dot
U_{\mu}=U^{\nu}U_{\mu;\nu}$, then the space-like heat flux density
four-vector is given by
\begin{equation}
Q^{\mu}=-\kappa h^{\mu\nu}(T_{,\nu}+TA_{\nu}).
\end{equation}
If the system concerned is in the thermal equilibrium, therefore the four-vector $Q^{\mu}=0$.\\
For the FRW background with thermal equilibrium condition satisfied,
we can have
\begin{equation}
\sigma_{\mu\nu}=0, \qquad  \theta=3H.
\end{equation}
And the Eq. (\ref{eq17}) reduces to
\begin{equation}
S^{\mu}_{;\mu}=\frac{\zeta}{T}\theta^2, \qquad S^{0}=\delta S,
\qquad S^{i}=0
\end{equation}
Hence, we get the differential equation for the entropy expression
with the argument as cosmic time below
\begin{equation}
S^{0}_{,0}+3H S^{0}=9\frac{\zeta}{T}H^{2}
\end{equation}
We therefore can convert the argument to  the scale factor $a(t)$,
by using the relation $\frac{\rd a}{\rd t}=aH$. Here we make an
assumption that the ambient temperature has the form that $T \propto
a^{-\gamma}$, or $Ta^{\gamma}=T_{0}$ with a free scaling parameter
$\gamma$, where $T_{0}$ is the temperature now:
\begin{equation}
\frac{\rd S^{0}}{\rd
a}+3\frac{S^{0}}{a}=9\frac{\zeta}{T_{0}}\frac{\dot a}{a}a^{\gamma-1}
\end{equation}
With already obtained relation as in the eq.(17):
\begin{equation}
\frac{\dot
a}{a}=\lambda_{2}a^{-1.5}-\frac{\lambda_{1}}{2n-3}a^{-n}+\frac{\lambda_{0}}{3}.
\end{equation}
and with the definition $\zeta=\frac{\lambda \rho_{cr0}}{H_{0}}$ and
$\lambda=\frac{\lambda_{0}}{9}+\frac{\lambda_{1}}{9}a^{-n}$, we
could write the differential equation as
\begin{equation}
\frac{\rd S^{0}}{\rd a}+3\frac{S^{0}}{a}=(\alpha + \beta
a^{-n})(\lambda_{2}a^{-1.5}-\frac{\lambda_{1}}{2n-3}a^{-n}+\frac{\lambda_{0}}{3})a^{\gamma-1},
\end{equation}
where the parameters $\alpha=\frac{\rho_{cr0}\lambda_{0}}{T_{0}}$ and $\beta=\frac{\rho_{cr0}\lambda_{1}}{T_{0}}$.\\
\begin{figure}
\includegraphics{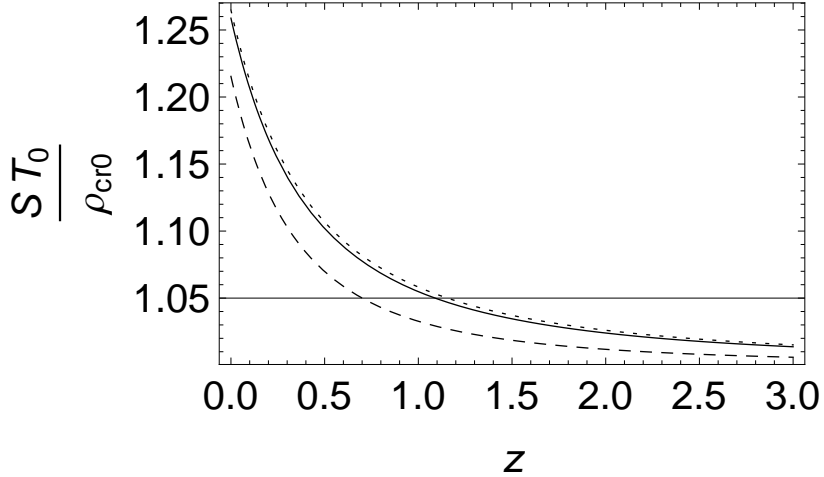}
\caption{The relation between the total entropy $aS^0$ of the
universe in the unified viscosity DE model and the corresponding
redshift. The dashed, solid and dotted lines correspond to the
temperature parameter $\gamma=0.5, 0.05$ and $0$, respectively. The
total entropy is increasing with the cosmic time flying (the
corresponding redshift becoming smaller), and this monotonous
increasing property can be regarded as an arrow of time pointing to
the universe evolution direction.}
\end{figure}
We work out the differential equation and find the general entropy
expression as below
\begin{eqnarray}
S^{0}& = & \frac{\alpha\lambda_{2}}{\gamma+1.5}a^{\gamma-1.5}-
\big[\frac{\beta\lambda_{0}}{3(\gamma-n+3)}-
\frac{\alpha\lambda_{1}}{(2n-3)(\gamma-n+3)}\big]a^{\gamma-n}+{}
                                                                \nonumber\\
& & {}+\frac{\alpha\lambda_{0}}{3(\gamma+3)}a^{\gamma}+
\frac{\beta}{\gamma-n+1.5}a^{\gamma-n-1.5}-{}
  \nonumber\\
& &
{}-\frac{\beta\lambda_{1}}{(2n-3)(\gamma-2n+3)}a^{\gamma-2n}+ca^{-3}{}.
\end{eqnarray}

For the special case with the $n=-1$, we can directly obtain
\begin{equation}
S^{0}_{n=-1}=\frac{\alpha\lambda_{2}}{\gamma+1.5}a^{\gamma-1.5}+
\big[\frac{\beta\lambda_{0}}{3(\gamma+4)}+
\frac{\alpha\lambda_{1}}{5(\gamma+4)}\big]a^{\gamma+1}+
\frac{\alpha\lambda_{0}}{3(3+\gamma)}a^{\gamma}+\frac{\beta}{\gamma+2.5}a^{\gamma-0.5}+
\frac{\beta\lambda_{1}}{5(\gamma+5)}a^{\gamma+2}+ca^{-3}.
\end{equation}

We plot the evolution trajectory of the total entropy calculated
above in the $S-z$ plane with dimension re-arranged. It is obvious
to see that the total system entropy is increasing with the cosmic
time flying, and this monotonous increasing property represents an
arrow of time for describing the universe evolution. Though the
quantum version of an arrow of time is still hotly debating the
thermodynamic entropy may be a good depiction to sketch the whole or
global universe evolution with very complicated ingredients inside.

\section{Summary and discussions}
In this letter we continue and largely extended our previous work on
the single and unified bulk viscous fluid as a potential dark energy
candidate by presenting an explicit viscosity form to mimic dark
energy behaviors and confront it with current observational data
sets. The specific feature here is a variable coefficient for the
new bulk viscosity form proposed, characterized by two free
parameters that can be best fitted by astrophysics observational
data sets. the best fitting results have shown that this concrete
model could yield theoretical prediction values in an acceptable
level by working out the numerical processing to the latest released
joint observational data sets.

Furthermore, we have performed the statefinder diagnostic parameter
analysis to this unified viscosity DE model, finding that in most
evolution stages of the unverse, statefinder parameters could be
used to obviously distinguish this viscosity DE model from the
$\Lambda$CDM model. But as shown in the figure of the statefinder
pair parameter \{$r,s$\}, the viscosity DE model evolution
trajectory passes the special point corresponding to the well known
$\Lambda$CDM model,
where the models degeneration emerges. The new $Om$ parameter
diagnostics made previously could be a more powerful tool then,
which might discriminate concrete models from the $\Lambda$CDM model
in the whole evolution history.

In this present work, we particularly concentrate on the singularity
behavior of the unified viscosity dark energy model. We find that
different parameter range selection, especially the region of power
parameter $n$, influences the finite future singularity. For the
$n=0$ case, which corresponds to the case with a constant viscosity,
there is an exact solution for the evolution scale factor. For the
early universe and the $n<0$ case, the solution of the scale factor
$a(t)$ has possessed the same structure. A further study of
viscosity effects on the early universe evolution, especially during
inflation stage, seems very interesting with rich possibilities.

In the context of the unified viscosity DE cosmology, we also
calculate the entropy for the total evolution universe, which has
been expressed in a complex form with this non-perfect viscosity
media. Calculations of the total entropy for the viscosity universe
evolution represent that the general second law of thermodynamics
holds in the whole or global universe description. The worked out
results show that in this complicated context, the total entropy is
increasing with the redshift decreasing, or cosmic time flying,
including the cosmic expansion accelerating stage as observational
data indicating now, which has been plotted in figure 5 with three
free parameter values chosen. The monotonous increasing property of
the total entropy with cosmic time flowing, the viscosity DE
universe evolution may provide us an arrow of time to describe the
complex universe changing direction. Though a definite concept for
the arrow of time is debating, now very puzzling in different
situations or versions: classical physics, quantum physics,
cosmology and quantum gravity, especially it is essential for a
correct quantum gravity theory to be expected to appear, the
thermodynamics second law is generally believed to hold to describe
the global evolution of our observational universe. So the
reasonable entropy expression is richly encoded helpful information
for the concerned system. It is certainly interesting and worthy of
further efforts.

Dark energy physics involves many fundamental concepts and beyond in
our already established "standard models" for both particle physics
and gravity. Viscosity media seems to relate the matter sector to
the geometric gravity side via the Einstein's general relativity
equation. Actually it also can be reconstructed effectively from the
left side to the right side of the equation by modified gravity or
extra dimension models for example, too, which is also intriguing.
With the available and upcoming high quality and increasingly large
amount of astrophysics observational data, especially the good low
redshift SNe Ia data sets with less uncertainty for possible errors
from the dust effects alleviated under control we expect the
ten-years old mysterious dark energy problem will be pinned down not
too long compared with the long time standing unsolved dark matter
mystery. Maybe the practically unified viscosity DE model or the
like can provide an economic mechanism to answer the both uncovered
secrets in one dark sector, a tale for the two mysteries.

\section*{Acknowledgement}
 We thank Profs. I. Brevik , M. Dabrowski, S.D.
Odintsov, Lewis H. Ryder for lots of interesting discussions during
the project. Also we thank Jie Ren for programming helps. This work
is partly supported by NSF of China under Grant No.10675062 and by
the project of knowledge Innovation Program (PKIP) of Chinese
Academy of Sciences under Grant No.KJCX2.YW.W10 through the KITPC
where we started this work.

\section*{Appendix}
We briefly summarize the joint statistical data analysis method here
by numerical processings to confront theoretical models with the
currently observational data sets. Supernova type 1a data provides
the direct evidence of cosmic accelerating expansion. Using the
$\chi^{2}$ statistics method, DE cosmological model parameters could
be best fitted. Recently, a large sample of low redshift($z<0.08$)
has been released freely\cite{data1}. Complied with the Union
sample\cite{data2}, the largest data set which could be obtained now
openly, is the Constitution data \cite{data3} with the newly high
quality low redshift data sets included. In this paper, we perform
the joint data fitting processings with this new compiled data set.
The observations of supernovas measure essentially the apparent
magnitude $m$, which is related to the luminosity distance $d_{L}$
by
\begin{equation}
m=M+5\log_{10}D_{L}\left(z\right),
\end{equation}
where the distance 
$D_{L}\left(z\right)\equiv\left(H_{0}\right)d_{L}\left(z\right)$ is
the dimensionless luminosity and
\begin{equation}
d_{L}=(1+z)d_{M}(z),
\end{equation}
where $d_{M}$ is the co-moving distance given by
\begin{equation}
d_{M}=\int_{0}^{z}\frac{1}{H(z^{'})}dz^{'}.
\end{equation}
Also,
\begin{equation}
\mathcal{M}=M+5\log_{10}\left(\frac{1/H_{0}}{1Mpc}\right)+25,
\end{equation}
where $M$ is the absolute magnitude which is believed to be constant
for all supernova of type Ia. The data points in these samples are
given in terms of the distance modulus

\begin {equation}
\mu_{obs}\equiv m(z)-M_{obs}(z).
\end{equation}
We employ it for doing the standard statistic analysis. So the
$\chi^{2}$ is calculated from
\begin{equation}
\chi^{2}=\sum_{i=1}^{n}\bigg[\frac{\mu_{obs}(z_{i})-\mu_{th}(z_{i};c_{\alpha})}{\sigma_{obs}(z_{i})}\bigg]^{2}.
\end{equation}
Theoretical calculation value is expressed as $\mu_{th}$, and
$\mu_{th}=M^{'}-5\log_{10}D_{Lth}(z_{i};c_{\alpha})$, where
$M^{'}=\mathcal{M}-M_{obs}$ is a free parameter need to fit as well
and $D_{Lth}(z_{i};c_{\alpha})$ is the theoretical prediction for
the dimensionless luminosity distance of a supernovae at a
particular distance, for a concreste model with some parameters
$c_{\alpha}$. The aid of this procedure is to determine the value of
$c_{\alpha}$.

On the other hand, the shift parameter $\mathcal{R}$ inferred from
CMB power spectrum of WMAP year five data and the distance parameter
$\mathcal{A}$ from the BAO data of LSS, like SDSS and 2dF, are
considered to give effective contributions to the joint statistical
data fitting. The shift parameter $\mathcal{R}$ is defined in
refs.\cite{A} and \cite{ap} as
\begin{equation}
\mathcal{R}\equiv\sqrt{\Omega_{m}}\int_{0}^{z_{*}}\:\frac{d
z^{'}}{h(z^{'})},
\end{equation}
and WMAP5 results \cite{w} have updated the redshift of
recombination to be at $z_{*}=1090$. Its detail meaning can be found
in reference \cite{a}. The distance parameter $\mathcal{A}$ is given
by
\begin{equation}
\mathcal{A}\equiv\sqrt{\Omega_{m}}\:h(z_{b})^{-\frac{1}{3}}\big(\frac{1}{z_{b}}\int^{z_{b}}_{0}\:\frac{d
z^{'}}{h(z^{'})})^{\frac{2}{3}},
\end{equation}
where commonly value $z_{b}=0.35$ is used.

Jointly taking into considering parameters $\mathcal{R}$ and
$\mathcal{A}$, we can soundly use the total $\chi^{2}_{total}$ to
make the best fitting analysis:
\begin{equation}
\chi^{2}_{total}=\chi^{2}+\left(\frac{\mathcal{R}-\mathcal{R}_{obs}}{\sigma_{\mathcal{R}}}\right)^{2}+\left(\frac{\mathcal{A}-\mathcal{A}_{obs}}{\sigma_{\mathcal{A}}}\right)^{2}.
\end{equation}
So far this is the most acceptable joint statistical data analysis
method for exploring the DE mystery and for the DE cosmology model
buildings to confront with the astrophysics observational data sets
obtained. Complementary with the increasing amount and accuracy of
lensing, cluster survey and others data sets to be obtained we are
confident it will be not far away to pin down the relative young
dark energy and long standing dark matter identities, and eventually
to solve the dark sector puzzles with  possible new discoveries.

\end{document}